\documentclass[useAMS,usenatbib]{mn2e}
\usepackage{hyperref}
\usepackage{graphicx}
 
\begin{document}

\title[A search for Planet 9 in the IRAS data]{
A search for Planet 9 in the IRAS data
}
\author[Rowan-Robinson M.]{Michael Rowan-Robinson$^{1}$\\
$^{1}$Astrophysics Group, Imperial College London, Blackett Laboratory, Prince Consort Road, London SW7 2AZ, UK\\
}
\maketitle
\begin{abstract}
I have carried out a search for Planet 9 in the IRAS data.  At
the distance range proposed for Planet 9, the signature would be a 60 $\mu$m unidentified
IRAS point source with an associated nearby source from the IRAS Reject File of sources
which received only a single hours-confirmed (HCON) detection. The confirmed source should be 
detected on the first two HCON passes, but not on the third, 
while the single HCON should be detected only on the third HCON.
I have examined the unidentified sources in three IRAS 60$\mu$m catalogues: some can 
be identified with 2MASS galaxies, Galactic sources or as cirrus. The remaining 
unidentified sources have been examined with the IRSA Scanpi tool to check for the signature 
missing HCONs, and for association with IRAS Reject File single HCONs.  No matches of interest
survive.
For a lower mass planet ($\le 5 M_E$) in the distance range 200-400 AU, we expect a pair or triplet of 
single HCONs with separations 2-35 arcmin. Several hundred candidate
associations are found and have been examined with Scanpi.  A single candidate for Planet 9
survives which satisfies the requirements for detected and non-detected HCON passes.  A fitted orbit suggest a 
distance of 225$\pm$15 AU and a mass of 3-5 $M_E$.  
Dynamical simulations are needed to explore whether the candidate is consistent with existing planet ephemerides. 
If so, a search in an annulus of radius
$2.5-4^o$ centred on the 1983 position at visible and near infrared wavelengths would be worthwhile.
\end{abstract}

\section{Introduction}
In the 1980s there had been a long history of interest in what would at that time have been
a tenth planet, Planet X.  There appeared to be unexplained residuals in the orbit of
Neptune.  Though these were much smaller than the Uranus residuals by which Le Verrier and Adams discovered
Neptune, they had motivated Tombaugh`s search for a new planet, which resulted in the
discovery in 1930 of what we now know as the dwarf planet Pluto.  It quickly became
clear that Pluto was too small to explain the Neptune residuals and so the possibility
of a tenth planet remained (see Batygin et al (2019) for a full historical account and references).

In 1983, while working on the preparation of the IRAS Point Source Catalog, I undertook
a systematic search for Planet X in the IRAS data.  The search was unsuccessful though it
did yield a detection of Comet Bowell (Walker and Rowan-Robinson 1984). Amusingly a
misunderstanding of an earlier IRAS science team briefing by senior NASA personnel resulted in a
press story in 1983 that IRAS had discovered Planet X (see Rowan-Robinson 2013 for full
details of how this misunderstanding arose).

Interest in Planet X flared up again in the late 1980s (Harrington 1988, Seidelmann and Harrington 1988, 
Jackson and Killen 1988, Neuhauser and Feitzinger 1991) and the Royal Astronomical
Society staged a discussion meeting in 1991 on `Solar System Dynamics and Planet X`. I gave a report
on my IRAS searches and concluded that I was 70$\%$ certain that Planet X 
did not exist.
The figure of 70$\%$ referred to the area of sky in which I was able to carry out my IRAS
search.  Reports of the meeting were given by Morrison (1992) and Crosswell (1991).

Subsequently a remeasurement of the mass of Neptune eliminated the Neptune residuals (Standish 1992).
The absence of deviations from the orbits of the Pioneer and Voyager spacecraft shows that
no unknown massive solar system planet resides in the plane of the ecliptic.  Luhman (2014) has used 
WISE data to set severe limits on Saturn or Jupiter mass objects in the solar system out to 
28,000  and 82,000 AU, respectively.

The discovery of dozens of new dwarf planets during the past twenty years has resulted
both in the redefinition of Pluto as a dwarf planet and in their potentiality as probes
of possible distant massive planets in highly inclined orbits.  Batygin and Brown (2016) and Brown and Batyin (2016),
developing an idea of Trujillo and Sheppard (2014),
proposed that a planet of a few tens of earth-masses in an inclined and eccentric
orbit at 280-1000 AU could explain alignments in the orbits of the Kuiper belt dwarf
planets.  As this was significantly more distant than the Planet X I was looking for in 1983,
I thought it worthwhile to redo my IRAS search and set out quantitatively what are the
constraints on such an object. Fienga et al (2016), Holman and Payne (2016), Iorio (2017), 
Millholland and Laughton (2017), Medvedev et al (2017), Caceres and Gomes (2018), Brown and Batygin (2019), 
Batygin et al (2019) and Fienga et al (2020), have given additional
dynamical constraints on Planet 9`s orbit.  Specifically Fienga et al (2016) argue against d $<$ 370AU
for a 10 $M_E$ planet, using Cassini radio ranging data, which they revise to d $<$ 650AU in Fienga et al (2020).  
 
Sections 2 discusses the IRAS capability to detect moving objects, section 3 focuses on
different categories of cool ($<$1000 K) movers, sections 4 reviews unidentified sources in the IRAS
PSCz and RIIFSCz surveys, section 5 discusses the IRAS 60 $\mu$m single HCON database.  Section 6 discusses a new
search for warm, slow movers, relevant to the search for Planet 9. Section 7 discusses how a Planet 9 candidate 
might be recovered and section 8 gives my discussion and conclusions.

\section{IRAS capability to detect movers}

IRAS surveyed 98 $\%$ of the sky at 12, 25, 60 and 100 $\mu$m.  The focal plane array
consisted of two strips of detectors for each of the four bands. The resolution of IRAS
at 60 $\mu$m was $\sim$1 arcmin.

The confirmation strategy for the IRAS Point Source Catalog (PSC: IRAS Explanatory Supplement, 1984) was at four levels:
(i) seconds confirmation, as sources crossed the two array strips for each waveband,
(ii) hours confirmation (HCON) as the same area was observed on two or sometimes three
successive orbits
(iii) weeks confirmation, as the same process was repeated roughly one to two weeks later (1$\%$ of the
sky received only a single HCON coverage)
(iv) months confirmation, when the process was repeated six months
later.  At the termination of the mission 75$\%$ of the sky had achieved a third HCON coverage.  
2 $\%$ of the sky had no coverage at all.  The IRAS coverage gaps, two
ribbons of sky centred at ecliptic longitudes 160$^o$ and 240$^o$, are edged by narrow strips 
in which only a single HCON was achieved.

This confirmation strategy allowed the identification of different categories of
moving object:

(a) fast movers:  Sources which passed seconds-confirmation but failed hours-confirmation
were examined in real time at the IRAS ground station at Rutherford Appleton Laboratory.
Several comets and Apollo asteroids were found and published (Davies et al 1984).  

(b) medium movers: sources which passed hours-confirmation, and so were recorded as HCONs, 
but failed weeks-confirmation were retained in a ‘PSC Reject File’.  These would include 
main-belt asteroids, which were published as the IRAS Asteroid and Comet Catalog (1986).  

For areas covered by 3 hours-confirming coverages we would expect to see 3 separate HCONs.

(c) slow movers: sources which passed weeks confirmation but failed months confirmation
would be present in the PSC but could have a separate single-HCON companion from the 
third coverage. These are the main cases of interest for a Planet 9 candidate.

(d) very slow movers: sources which pass months confirmation will be present in the PSC.
Separate analysis of time-line data would be needed to reconstruct motions of moving
sources which moved less than $\sim$ 1 arcmin in 6 months.  However such objects would have to be at distances
$>$ 7000 AU and so are not of interest as Planet 9 candidates.

What positional disagreement results in non-confirmation of HCONs ?  This varies with 
direction on the sky but the 60 $\mu$m (2-$\sigma$) error ellipses for confirmed sources have semi-axes typically 
30 arcsec (cross-scan)
by 10 arcsec (in-scan).  I adopt 1 arcmin as the characteristic motion which could cause
confirmation failure at 60 $\mu$m.

\begin{figure}
\includegraphics[width=9cm]{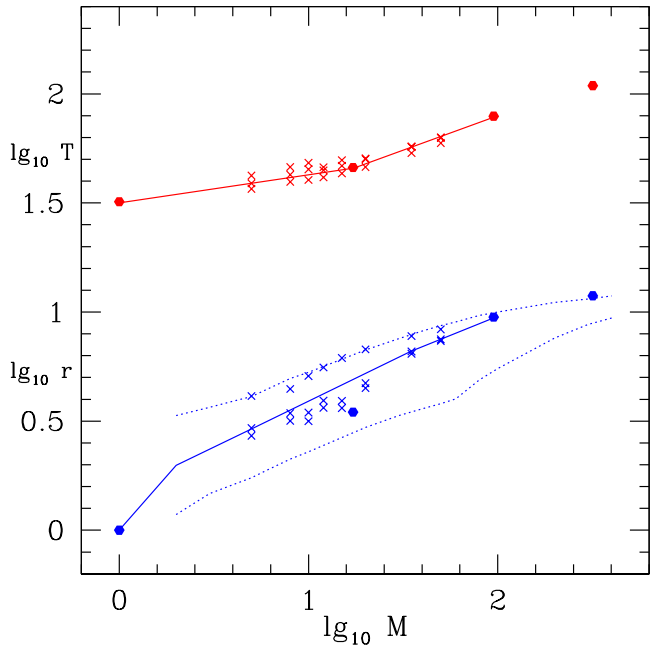}
\caption{
$log_{10}$ planetary radius (in earth radii, blue), and $log_{10}$ predicted temperature ($^o$C, red) due to 
internal heating only, versus $log_{10}$ planetary mass (in earth masses).
Filled red hexagrams are temperature predictions from Reynold et al (1980), while filled blue hexagrams are 
data for solar system planets.  
Crosses are for recent Planet 9 models explored by Fortney et al (2016, their Table 1). Blue dotted curves indicate range of predicted radii 
for exoplanet models of Mordasini et al 2012 (their Fig 10).  Solid lines indicate relations assumed for predictions of Fig 2. 
}
\end{figure}

\section{IRAS sensitivity to cool movers}
The capability of IRAS to detect cool ($<$ 1000 K) nearby objects was discussed by Reynolds,
Tarter and Walker (1980), who compared the expected sensitivity of IRAS to existing
(partial) ground-based surveys.  They gave the expected surface temperature of the known
planets (and hypothetical objects of mass greater than Jupiter) if they were located
at great distances from the Sun, and therefore radiating internal heat only.  
Figure 1 (upper part) shows their estimates of surface temperature for different masses and also the 
temperatures calculated by Fortney et al (2016) in their recent detailed study of possible Planet 9 models.
My adopted mean relation is shown as a solid line.  The lower part of Fig 1 shows the radii of the
solar system planets, the model estimates of Fortney et al (2016), and also the range of model radii considered
by Mordasini et al (2012) in their discussion of possible exoplanet models, again with my
adopted mean line.  We can then construct a mass-distance sensitivity diagram (Fig 2)
which shows the distance to which different mass planets would be detectable in the four IRAS bands.  
The prime search-area for a Batygin and Brown object would be 5-20 $M_E$, with the distance range 
being 280-1100 AU.  
At 60$\mu$m IRAS would 
in principle have been able to detect a 5 $M_E$ planet to 220 AU, a 10 $M_E$ planet to 380 AU, and a 20 $M_E$ planet to 690 AU. 
But if we allow for the maximum radii considered by Mordasini et al (2012), 
a 5 $M_E$ planet could be detected to 350 AU, a 10 $M_E$ planet to 600 AU, and a 20 $M_E$ planet to 1000 AU. These are worthwhile limits.

\begin{figure*}
\includegraphics[width=14cm]{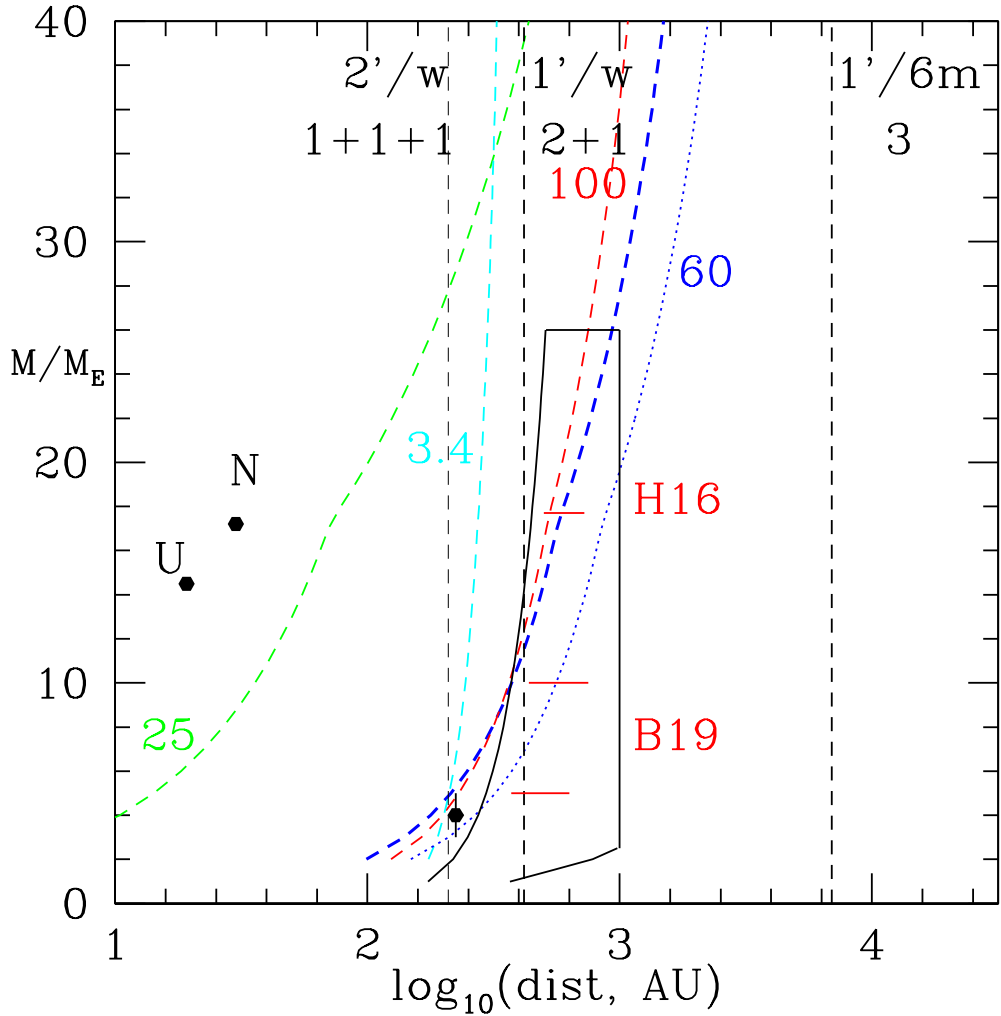}
\caption{
Planetary mass (in earth units) versus distance (in AU).  Green, blue and red dashed loci show limit of IRAS
25, 60 and 100 $\mu$m detectability (assumed flux-limits 0.25, 0.45 and 1.0 Jy).  The right-hand blue dotted locus corresponds to the extended planetary models of 
Mordasani et al (2012).  Vertical dotted lines divide distances into those in which three
separate single HCONs would be seen (1+1+1), those at which sources would weeks-confirm but fail 6-months
confirmation (2+1), and those beyond which all three HCONs should be combined together (3).  The black locus 
contains the region in which a planet of 1-26 earth-masses, at 180-1000 AU could be found:  the left-hand boundary
is the limit set by Cassini ephemerides, the right-hand boundary is the limit for non-disruption during planet formation, 
objects below the lower right boundary would be dwarf planets (Batygin et al 2019).  U, N indicate the locations of Uranus 
and Neptune.  The cyan locus is the limit for the WISE search at 3.4$\mu$m 
(Meisner et al 2018),which covers 70$\%$ of the sky to a depth of m(3.4$\mu$m)=16.7 (Vega), for an assumed albedo 0.75.  
Red horizontal bars denote the preferred mass-distance ranges of Holman (2016) and Brown and Batygin (2019).  The Planet 9 candidate discussed in sections 6.5 and 7 is shown as a black hexagon with error bars.
}
\end{figure*}

The main motion of distant solar system movers is an ellipse due to parallax, with major diameter
2$\alpha(arc min)$, where $\alpha=3438/d(AU)$.  I have shown a vertical dashed line in Fig 2 corresponding to
$\alpha=0.5arcmin$, essentially the most distant solar object for which IRAS could detect motion (1arcmin 
in 6 months).  Two
weeks-confirming HCONs would have a separation of 1 arcmin when $\pi \alpha(arc min)=26$, or d$\sim$ 415AU
and this is also shown (thicker dashed line). For the proposed Planet 9 at 280-1100 AU, the expected parallactic motion 
in 6 months would be 6-24 arcmin, easily detectable by IRAS.  Superposed on the elliptical motion on the sky due to 
parallax there would be a drift along a great circle due to the orbital motion, typically 0.3-1.3 arcmin in
6 months for d=1100-400AU.


Fienga et al (2016) have set a dynamic constraint of d$>$370 AU for a 10 $M_E$ planet on a particular
hypothetical Planet 9 orbit, based on the ephemeris of Saturn as observed by Cassini.  Batygin et al
(2019) have scaled this as $M/r^3$ to give the left-hand edge of the permitted region in Fig 2.

Fortney et al (2016) and Meisner et al (2018) have discussed the detectability of Planet 9 in the 
WISE W1 (3.4$\mu$m) channel.  We have shown the corresponding locus for an assumed WISE 3.4$\mu$m
sensitivity of 16.7 m(Vega), which Meisner et al (2018) have achieved for 70$\%$ of the sky.  
At 3.4$\mu$m the emission from Planet 9 will be through a combination of reflected light from the Sun, which depends on the
albedo at 3.4$\mu$m, and emission.  The net flux is highly dependent on assumptions about the Planet 9 atmosphere and 
abundances, especially of methane (Fortney et al 2016).  We have adopted the Rayleigh scattering limit albedo
of 0.75, though for most Planet 9 models, and for Neptune, the albedo would be orders of magnitude 
smaller than this.  Even with this optimistic assumption and the adopted mean radius-mass line of Fig 1, the
predicted 3.4 $\mu$m flux from reflected light for Planet 9 in the expected mass-distance range is not really detectable by WISE.
However the most extreme model of Fortney et al (2016) does give such strong molecular emission that the 
effective albedo becomes $>$ 1 and such model planets would be detectable by WISE at 3.4$\mu$m.
The range of radius predictions at a given planet mass makes the limiting distance estimates uncertain by
$\pm 50\%$.



Unfortunately there are a number of alibis for Planet 9: (i) IRAS has no coverage at all in 
about 2$\%$ of the sky and only a single coverage in a further 1$\%$, (ii) some IRAS galaxy catalogues (e.g. PSCz, QIGC) 
impose additional constraints, excluding regions of the sky
with high source-density flag at 12, 25, or 60 $\mu$m (PSCz, QIGC), with high-source-density flag at 100 $\mu$m (QIGC), or with 
high extinction (PSCz).These constraints, which have a similar effect, exclude a further $\sim 13 \%$ of the sky.
(iii) a quarter of the sky received only
2 hours-confirmed coverages, which makes detection of slow movers difficult, unless
they are detected in 3 bands so that colour information can be used.


\section{Unidentified sources in the IRAS surveys}
Our starting point in any search for Planet 9 in the IRAS data is the unidentified sources in the IRAS Point source Catalog (PSC) 
and Faint Source Catalog (FSC) surveys.  In this section I discuss the unidentified sources in the PSCz catalogue (Saunders et al 2000),
 as the most complete and well-studied compilation from the IRAS PSC survey, and in the RIIFSCz galaxy catalogue (Wang et al 2014),
 as the best studied compilation from the IRAS FSC survey.

\subsection {Unidentified sources in PSCz}
For slow movers we are interested in PSCz sources which are not identified with galaxies or Galactic objects.
The IRAS PSCz survey sought to identify and get optical spectra of every
IRAS PSC 60$\mu$m sources brighter than 0.6Jy, after exclusion of Galactic sources and cirrus. The full catalogue consists of 
15459 sources, of which 15200 have $|b| > 5^o$.
Of the latter all but 791 (5.2$\%$) have measured redshifts (these numbers differ slightly from those given by Saunders et al 2000). 
$[b|<5^o$ sources are discussed below. The PSCz completeness limit is given as 98.5$\%$ outside the masked area (see Fig 5), ie for 85$\%$
of the sky.

I have examined these 791 sources in the NASA$/$IPAC Extragalactic Database (NED). 493 were either identified with catalogued galaxies 
(especially with 2MASS galaxies), were identified galaxies in the RIFSCz catalogue (Wang et al 2014), were classified
as cirrus in the literature, were classified as known Galactic sources, or were within 1.5 arcmin of a 2MASS galaxy.
A plot of 60$\mu$m flux versus K-band magnitude for IRAS Faint Source Catalogue galaxies shows that
most, but not quite all, IRAS galaxies with 60 $\mu$m flux density $>$ 0.6 Jy are likely to be detected as 2MASS sources. 

The distribution of distances of the nearest 2MASS galaxy for sources not associated with 2MASS galaxies
in NED is consistent with being the tail of a 30 arcsec Gaussian, which is a plausible value for the rms error in IRAS-2MASS
associations. Most sources with separations less 
than 1.5 arc min are therefore likely to be real associations and few of those with larger separations are likely to be
true associations.  However because the size of the 60 $\mu$m detectors (4.5x0.5 arc min), errors in processing 
can occasionally result in separations between an IRAS source and its associated galaxy of more than 1.5 arc min.


304 (2$\%$) sources remain unidentified in NED.  Many of these are at $|b| = 5-20^o$, where identification is harder 
because of strong extinction in the optical and because of the high surface-density of stars, and where the probability of
spurious cirrus sources is higher.  
Reviews of targeted searches for low latitude galaxies have been given by Kraan-Korteweg and Lahav (2000) 
 and Staveley-Smith et al (2016).
Examination of the Schlegel et al extinction maps (Schlegel et al 1998) suggests that a further 72 of these 304 
unidentified sources are cirrus.
164 are classified as g or gf (‘faint galaxy’) in the NASA-HEASARC PSCz archive, although without confirmation of the 
source position by radio observations (available for only 27 of the sources)
such an identification is ambiguous, since several faint objects lie within the IRAS error ellipse. For a further
three, a radio association confirms the source as a galaxy.
HEASARC identify a further 2 objects as stellar and 8 as cirrus.
3 are sources recovered by Saunders et al (2000) from the IRAS Reject File, but without a confirming galaxy or
IRAS FSC association the reality of these sources is doubtful.  The final number of unidentified IRAS PSCz sources
is then 52, if ‘gf’ designation is accepted as an identification, and IRAS Reject File sources are rejected.
However the resulting constraint on very slow movers brighter than 0.6 Jy at 60 $\mu$m is still rather limited, that 
the surface-density is no greater than 0.001 per sq deg.  But from Fig 2 it is clear that no very slow movers 
 of interest (parallactic angle $> 3 arcsec$, $M < 100 M_E$) would
have 60 $\mu$m flux-densities greater than 0.6 Jy. Unidentified PSCz sources could still be of interest as candidates 
for Planet 9 if they lay within the areas that received only 2 HCON coverages.  

\subsection{Unidentified sources in the RIIFSCz catalogue}
The revised Imperial IRAS FSC redshift (RIIFSCz) catalogue (Wang et al 2014) is also of interest to look for unidentified
60 $\mu$m sources to fainter fluxes than the PSCz catalogue.  The IRAS Faint Source Catalog was processed in a different way to the PSC,
essentially coadding all the IRAS data before searching for point sources.   
Moshir et al (2019) give the 90$\%$ completeness limit for the IRAS Faint Source Catalog at 60 $\mu$m as 0.28 Jy and this is confirmed
by 60 $\mu$m source-counts.
There are 2727 sources in the RIIFSCz brighter than 0.3 Jy which do not have a galaxy association.  I have examined all of these 
using the National Extragalactic Database (NED) and find that 1720 have plausible associations with 2MASS galaxies, which
I define as follows:  for ‘bright’ 2MASS galaxies (K$<$13.5) an association within 1.5 arcmin; for 2MASS galaxies with 
13.5$<$K$<$14.5 an association within 1.0 arcmin; and for ‘faint’ 2MASS galaxies (14.5$<$K$<$15.5) an association within
30 arcsec.  The redshifts of the associated galaxies, where available, are generally in the range 0.01-0.15.  
A few are associated with stars (68) or other Galactic sources (2).  The remaining 937 sources, 3.1$\%$ of sources brighter 
than 0.3Jy, are unidentified and may be higher redshift(z$>$0.1) galaxies or spurious sources. I investigate whether any of
these can be associated with a single HCON source below (section 6.2).

The distribution on the sky of PSCz and RIIFSCz unidentified sources is shown in Fig 5.


\section{The IRAS 60 $\mu$m single HCON database}
To find a Planet 9 candidate in the IRAS data we need to look for an unidentified point source associated with a
separated single HCON source.  Here I characterise in detail for the first time the IRAS single HCON database 
(Point Source Reject File).  This consists of 21045 single HCONs with detections at 60 $\mu$m.  For one
reason or another these have failed to be associated with IRAS PSC sources.  These can include
main-belt asteroids, planets, distant comets, cirrus, galaxies whose different HCONs did not get properly processed
together, and point sources in areas of the sky which achieved only a single HCON coverage.
Counts of 60$\mu$m single HCONs 
suggest the 90$\%$ completeness limit is 0.45 Jy.

12760 of these have detections also at 100 $\mu$m 
 and for these Fig 3R shows S100 versus S60.  For comparison a similar plot is shown (Fig 3L) for
the QMW IRAS Galaxy Catalogue (Rowan-Robinson et al 1991).  This contains 17706 galaxies and differs from PSCz in 
including sources with 60 $\mu$m fluxes in the range 0.5-0.6 Jy).  The straight lines which bound most IRAS galaxies
correspond to $log_{10} S100/S60$ = 0.7 and -0.1.  Strong concentrations of single HCON sources
lie outside the region occupied by galaxies.  I divide the single HCONs into (i) hot sources 
( $log_{10} S100/S60 < -0.1$), (ii) warm sources ($-0.1 <  log_{10} S100/S60 < 0.7$), (iii) cold
sources ( $log_{10} S100/S60 > 0.7$.  Figure 4 shows the sky distribution, RA versus dec, for the three
classes, with the loci of the ecliptic and Galactic planes indicated.  


It is apparent that most of the hot single HCONs are main-belt asteroids and that most of the cold
single HCONs are Galactic cirrus.  The warm sources show concentrations along the edges of the IRAS
coverage gaps, with a strong concentrations where these gaps cross the Galactic plane.  Otherwise
the distribution of warm single HCONs is broadly isotropic, consistent with most being incorrectly
processed bits of galaxies. 

\begin{figure*}
\includegraphics[width=7cm]{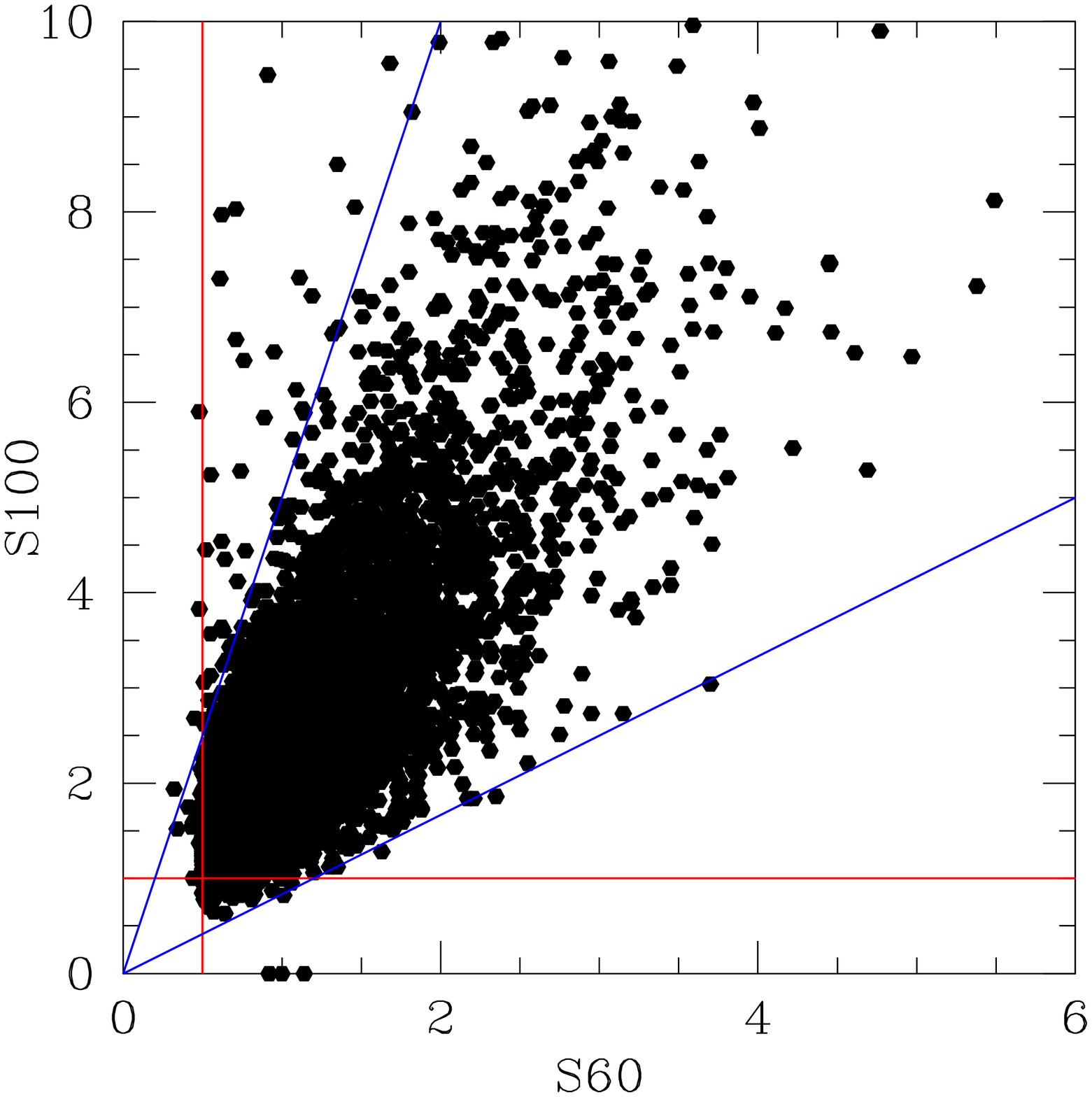}
\includegraphics[width=7cm]{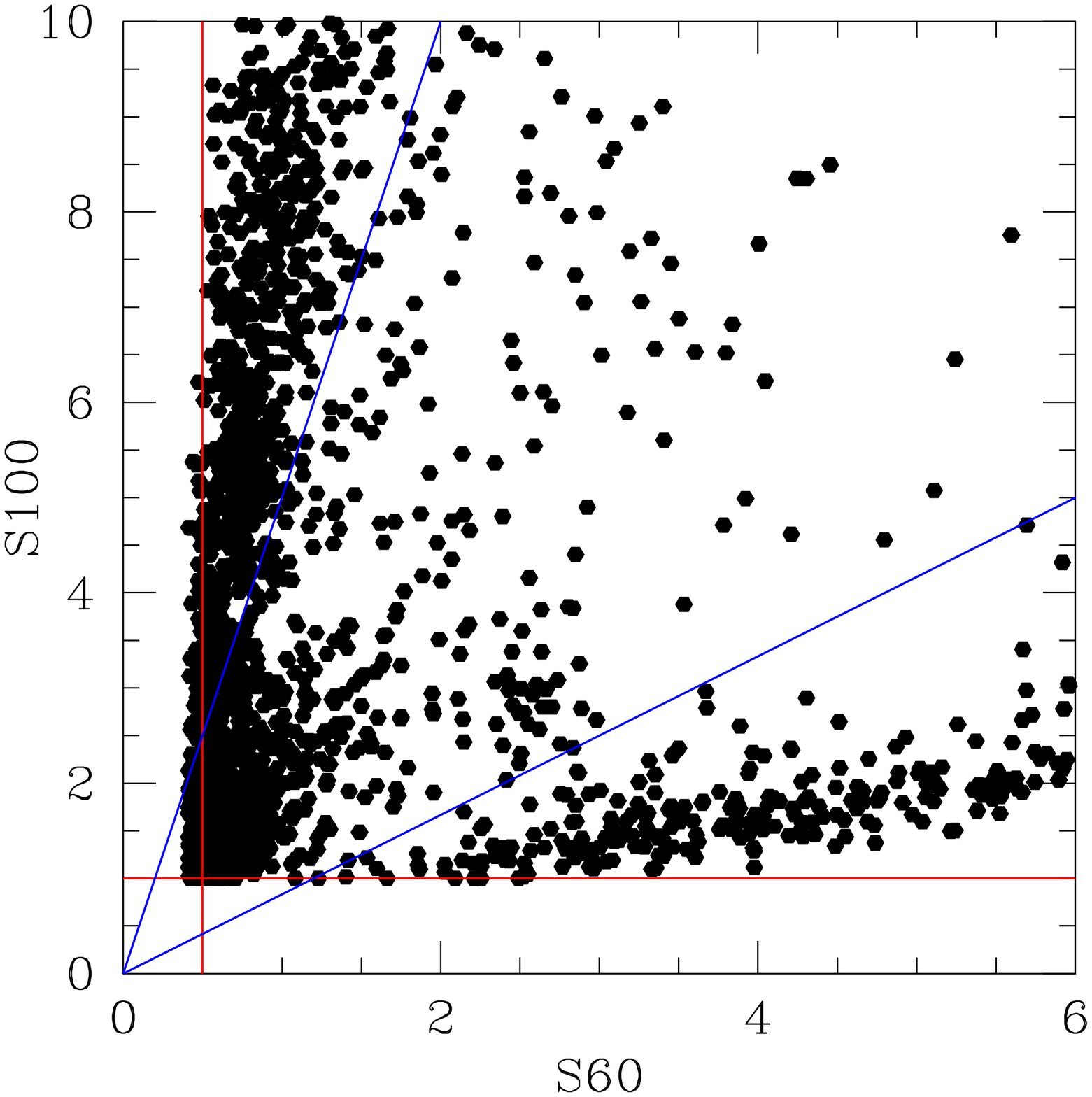}
\caption{
L: S100 versus S60 (Jy) for sources from the QMW IRAS Galaxy Catalogue (Rowan-Robinson et al 1991).
R: S100 versus S60 (Jy) for isolated single HCON sources from IRAS Reject File. The blue lines
demarcate ‘hot movers’, S100$/$S60$<$0.8 (T$>$82K), ‘warm movers’ $0.8<S100/S60<5$ (30$<$T$<$82 K), 
and ‘cold movers’, $S100/S60>5$ (T$<$30K).
}
\end{figure*}

\begin{figure*}
\includegraphics[width=5cm]{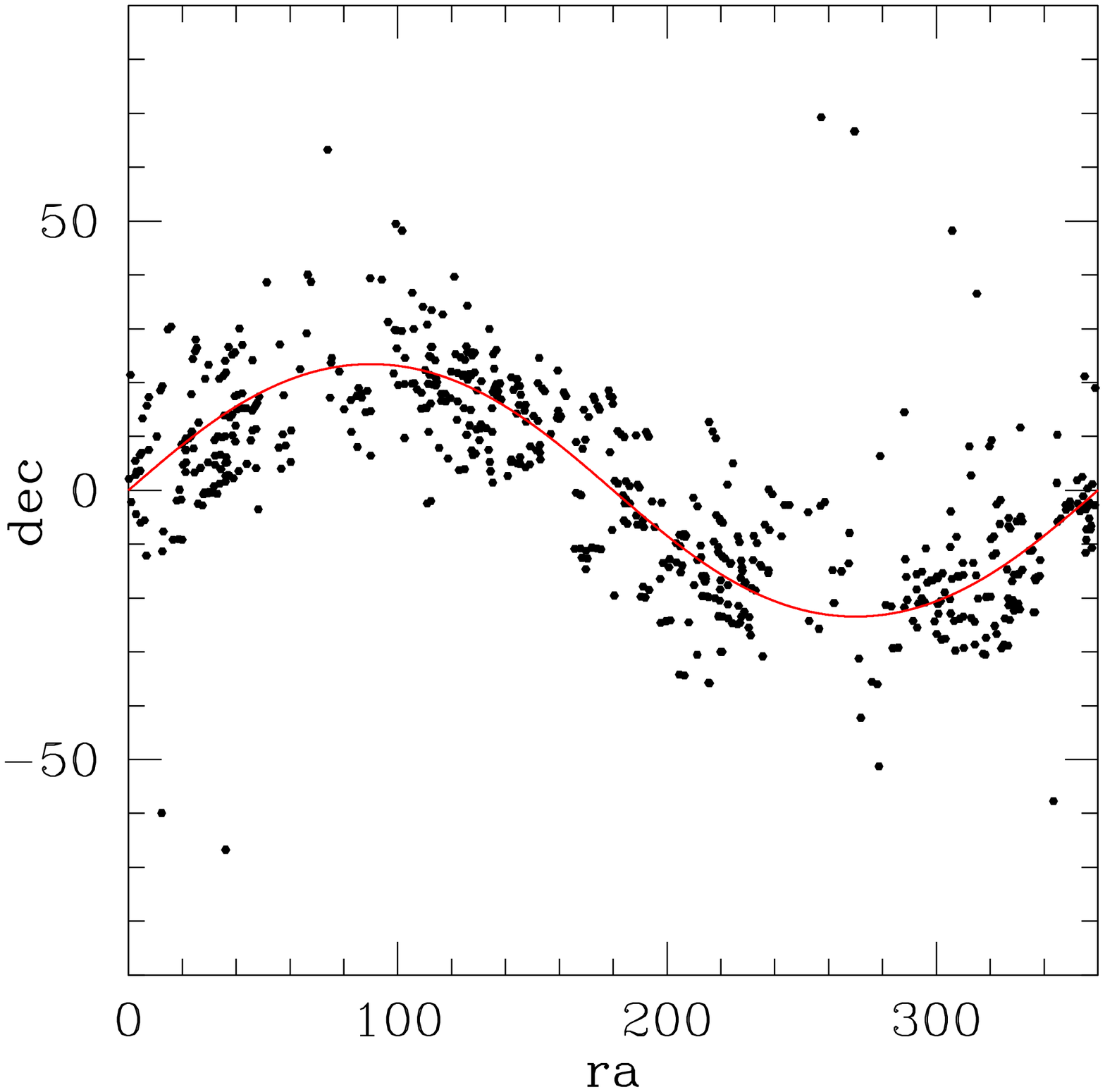}
\includegraphics[width=5cm]{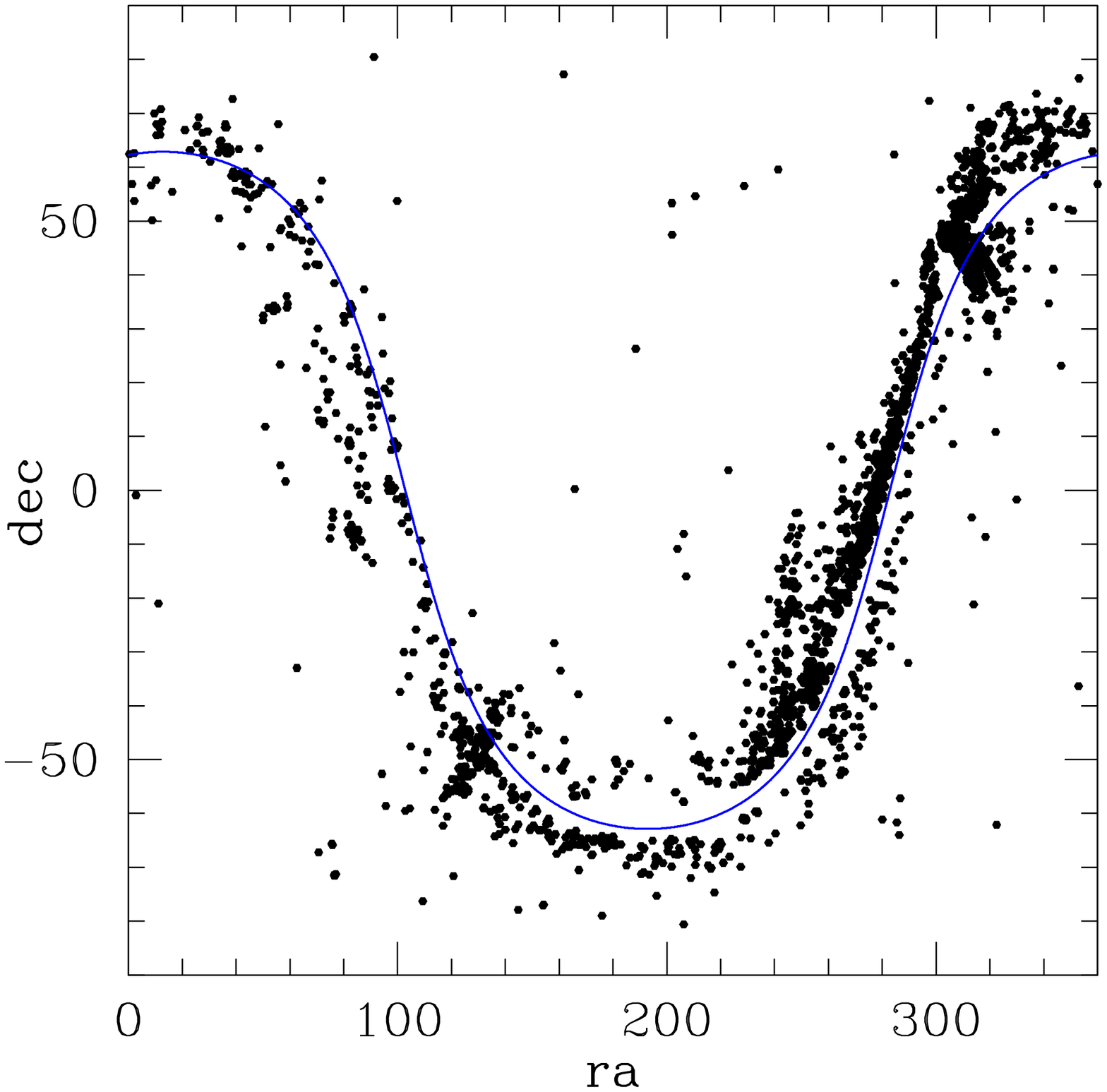}
\includegraphics[width=5cm]{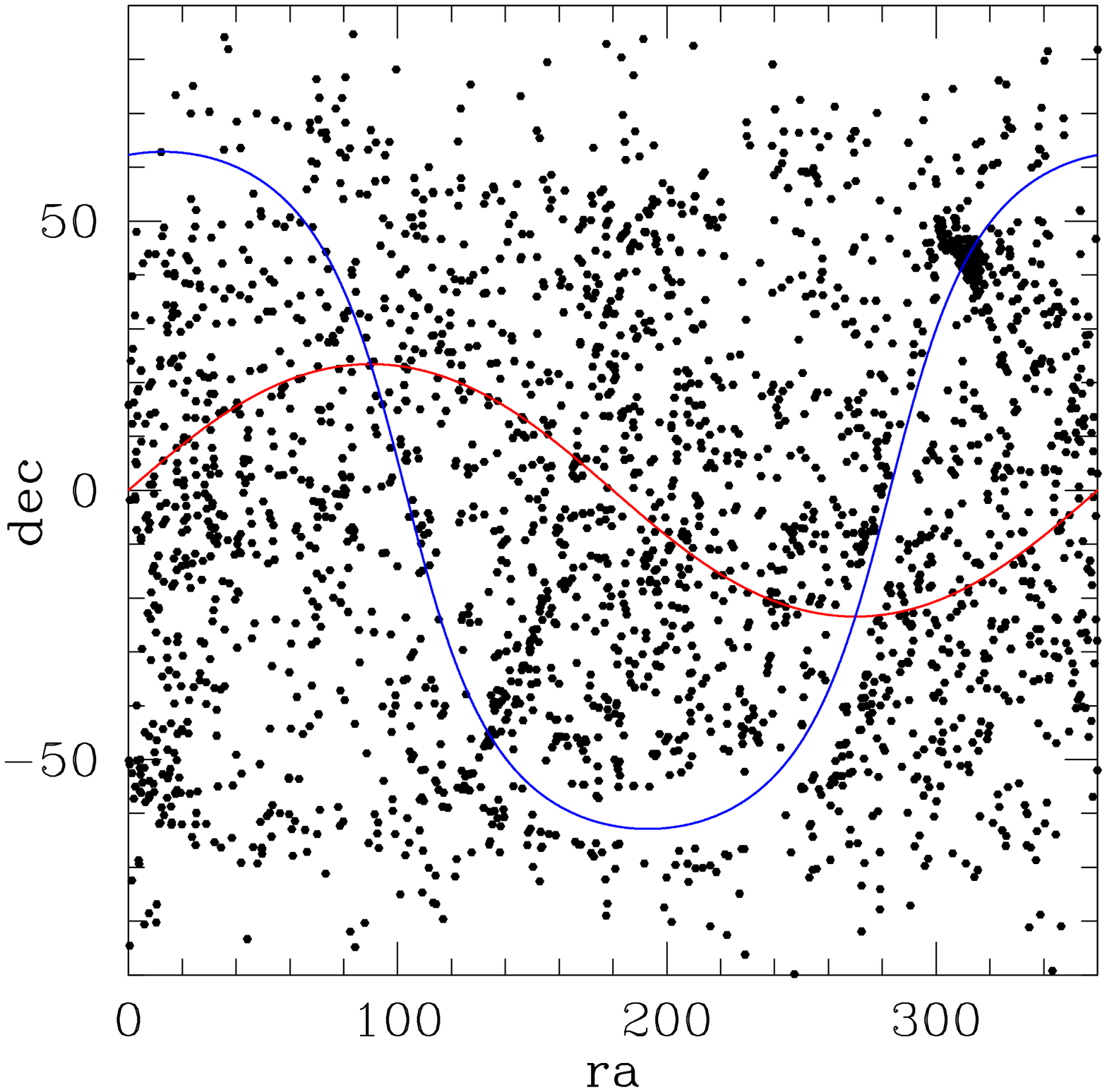}
\caption{
L: RA versus dec for hot single HCONs. Most are confined to a band around the ecliptic plane and are
main belt asteroids.
centre: RA versus dec for cold single HCONs. Most lie close to the Galactic Plane and are due to cirrus.
R: RA versus dec for warm single HCONs.  Apart from heavy concentrations where the IRAS coverage gaps meet
the Galactic Plane, and a weaker concentration along the IRAS coverage gap, the distribution is relatively isotropic, 
indicating that most are probably galaxies.
}
\end{figure*}


\section{New search for warm slow movers (Planet 9 candidates)}
We now look for Planet 9 candidates.  Firstly I look for associations of unidentified sources from the different IRAS catalogues
with single HCONs, which would be candidates for Planet 9 at distances $>$ 400 AU.  Then I look for cross-associations of single HCONs, which would be candidates for a lower distance Planet 9.

\subsection{Association of PSCz unidentified sources with single HCONs}
To see whether there any candidates in the IRAS database for the hypothetical Planet 9, an object of a few tens of
earth-masses at a distance of 280-1100 AU, I have first searched for 60 $\mu$m-detected warm single HCONs 
(including 60 $\mu$m only objects) at distances of
2-25 arcmin from the 52 unidentified PSCz sources at $|b|>5^o$.  8 sources found 1HCON matches.  There were 
two cases where multiple single HCONs
were found associated with the same PSCz source.  These are presumed to be either main-belt asteroids or due
to cirrus or other Galactic structures.  Each of the remaining associated single HCONs was examined for
associations in NED.  Associations of 1HCON sources with 2MASS galaxies were accepted up to 2 arcmin, allowing for the poorer
error ellipse for single HCONs.  A number
of single HCONs were identified as due to cirrus from the Schlegel et al extinction maps.  

One further constraint is that the PSCz 
source should be missing the third HCON.  I have examined all 52 unidentified PSCz sources using the IRSA Scanpi tool, which allows the individual IRAS scans crossing any specified
position to be examined.  The signature of a Planet 9 candidate would be a source seen on the scans associated with the first two
HCONs but missing on the third HCON. One source, 88.313751+45.864166, showed this signature, and has a single HCON nearby 
(87.50875+45.64139, at 36.3 arcmin), but the latter is detected on all three HCONs and is not a mover. Also the flux agreement 
is poor (0.8Jy for PSCz source, 0.4Jy for 1HCON), so this source is not a convincing candidate for Planet 9.

Of the 259 PSCz sources at $|b|<5^o$, 105 did not have redshifts determined by Saunders et al (2000).  I
examined these for associations in NED and found 45 did not have associations.  These 45 were then 
run with Scanpi.  2 were spurious, 4 were cirrus, 2 probably Galactic, and there were just two with the 
characteristic missing HCON profile: 74.024582+47.587219 and 85.358749+21.475834. The first does not have 
a nearby single HCON of interest (the nearest is at 36 arcmin and is not unidentified).  The second has a 
nearby HCON (see Table 1) but this does not show any convincing detections on the Scanpi scans.

The conclusion of the search in PSCz is that no plausible candidate survives.

\begin{figure*}
\includegraphics[width=14cm]{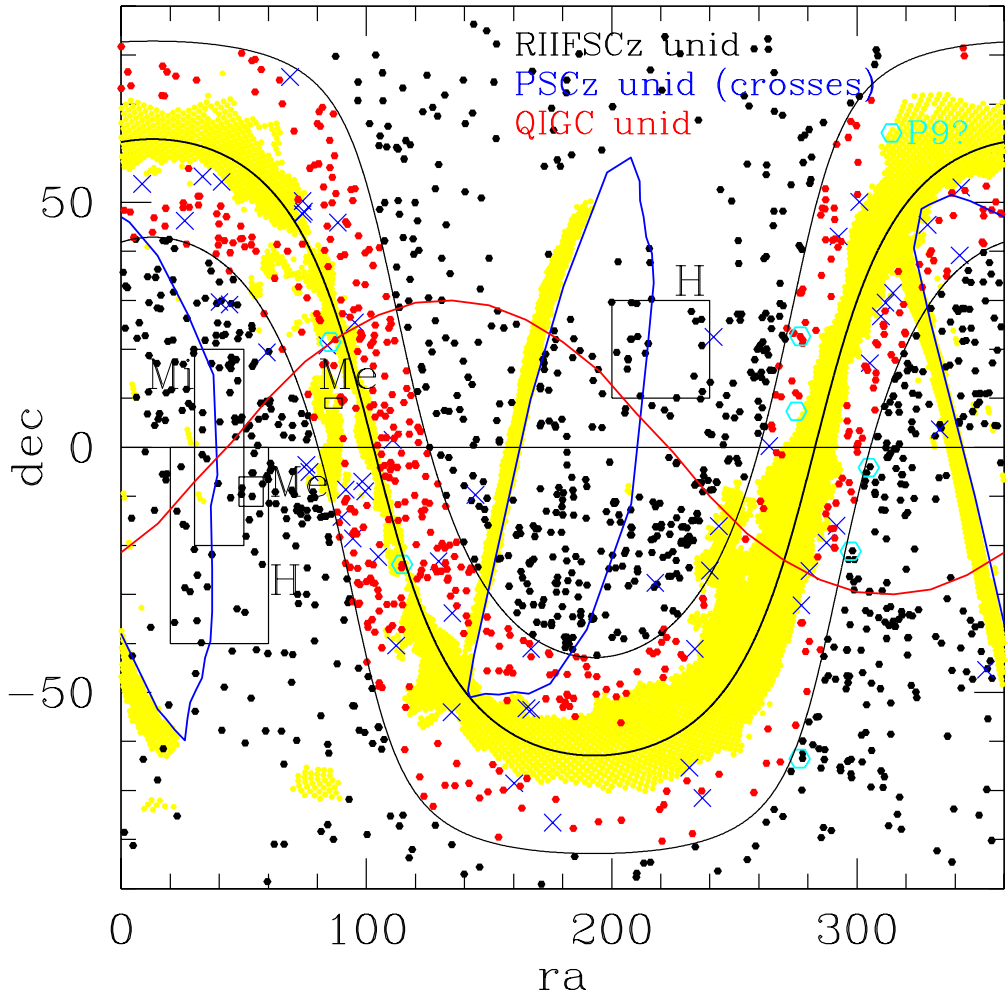}
\caption{
RA versus dec for remaining unidentified PSCz sources (blue crosses), RIIFSCz unidentified sources (black filled hexagrams), 
and QIGC unidentified sources at $|b|< 20^o$ (red filled hexagrams).  Areas which did not receive a third HCON coverage are delineated
in blue. The IRAS PSCz mask (zone of avoidance) is indicated in yellow. The areas in which it has been proposed that Planet 9 may lie are indicated as black rectangles
(H: Holman and Payne 2016, Me: Medvedev et al 2017, Mi: Millholland and Laughton 2017).  Galactic latitudes 0 and $\pm 20^0$ are shown as black loci 
and the red locus is the preferred Planet 9 orbit of Batygin et al (2019).  My Table 1 sources of interest are indicated 
as cyan open hexagrams.
}
\end{figure*}

\subsection{Association of RIIFSCz unidentified sources with single HCONs}
I looked for associations of single HCONs with the 937 unidentified RIIFSCz sources, at separations between 2 and 25 arcmin.
A total of 126 associations were found. The associated RIIFSCz sources were checked with Scanpi, looking for the pattern of
detections on the first two HCONs and non-detections on the third.  The single HCONs were examined with NED, so that identified
sources could be rejected, and with Scanpi to check the reality of the sources.

There were three cases cases of interest ( listed in Table 1).  However for two, the single HCON is not unidentified.  
The third RIIFSCz source is non really undetected on the third HCON and the flux match with the single HCON is poor.

The conclusion is that no Planet 9 candidate survives in the RIIFSCz catalogue.

\subsection{Association of QMW IRAS galaxy catalogue unidentified sources with single HCONs}
While the RIIFSCz takes our search down to 0.3 Jy, it does so only for $|b|>20^o$.  At $|b|<20^o$ we can use the QMW
IRAS Galaxy Catalogue (QIGC: Rowan-Robinson et al 1991) to extend the search to $\sim$0.5 Jy. The completeness of
QIGC is given as 80$\%$ in the flux-range 0.5-0.6Jy (Rowan-Robinson et al 1991).
I therefore searched for unidentified IRAS PSC sources with 60 $\mu$m fluxes in the range 0.5-0.6 Jy (from the QMW
IRAS Galaxy Catalogue, see Fig 5), with single HCON associations at separations in the range of
interest for Planet 9. 90 sources have 1HCON associations,
14 are unidentified: of these, 6 are spurious (Scanpi), 2 are extended at 60 $\mu$m, 7 were candidates for having missing HCONs, and
2 of these are of interest (Table 1).  For 114.505836-23.892778 the QIGC source and the single HCON have the appropriate pattern 
of detected and non-detected HCONs for a mover, in terms of 3-$\sigma$ detections in Scanpi.  However the non-detected HCONs are not 
very convincing because each has several 2-$\sigma$ detections.  For 276.900848+22.634167 the single HCON is not unidentified.

\begin{figure}
\includegraphics[width=7.0cm]{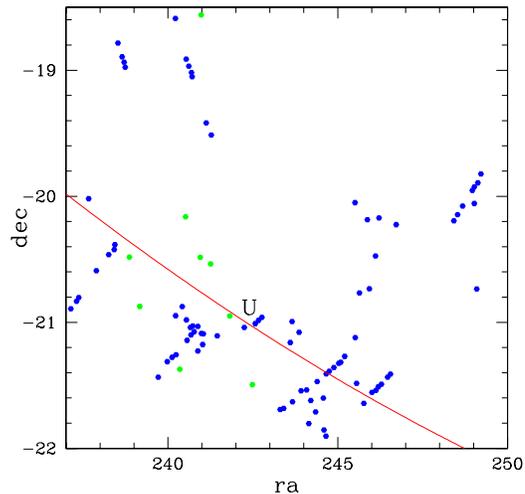}
\caption{
Distribution of 60 $\mu$m single HCON sources close to ecliptic plane (shown as a red locus). At least 12 main-belt
asteroid tracks can be seen. The triplet of points labelled U are due to Uranus.
}
\end{figure}

\subsection{Cross associations of single HCONs}
If Planet 9 lay at a distance 200 - 400 AU, it could be present in the IRAS Reject File of single HCONs as a triple 
of single HCONs with two separated by a few arc minutes and a third at 10-35 arc mins. The signature would be that 
the source would only be detected on one HCON pass each time, i.e. there must be non-detections as well as detections.  
In the areas which received only 2 HCONs there could be two single HCONs separated by a few arc minutes (see below).
Finally in the case of a faint planet 9 candidate at 280-1000 AU in the 3 HCON sky there could be two single HCONs 
separated by 6-25 arc minutes if the source failed one of its first two HCONs. For completeness I have explored all 
of these possibilities.

There are 15162 warm 60 $\mu$m single HCONs (including those undetected at 100 $\mu$m).  Many are spurious detections (Scanpi shows no pass
with detection $>3\sigma$), or are clearly extended, due to cirrus.  To cut down on the large number of uninteresting
matches with cirrus, extended or spurious sources, I require the 100 $\mu$m flux or upper limit to be $<$ 3.5Jy.  This
reduces the probability of a chance association with a single HCON within 25 arcmin at any given position on the sky to 7$\%$.

Some single HCONs are main-belt asteroids.  From Fig 3 these 
would be fainter (and smaller) asteroids, with S60 $<$ 3 Jy, otherwise they would be detected at 100 $\mu$m
and excluded as ‘hot’ movers. Fig 6 shows the distribution of single HCONs in a 40 sq deg area close to the ecliptic plane 
in which the tracks of at least 12 main-belt asteroids can be seen.  The triplet of sources labelled U are due to Uranus.  The
central source is Uranus itself. The two much fainter flanking sources appear to be glints from Uranus (they
are too far from Uranus to be the latter`s brighter moons).  In fact of the 88 single HCONs in this area of the sky as many
as 74 can be attributed to asteroids.  
These detections of Uranus and asteroids demonstrate the IRAS capability to detect faint moving objects.

There are 532 pairs of single HCONs at separations 2-25 arcmin.  I examined all of these with Scanpi and found one 
possible case of interest, 275.30018+7.33028 (Table 1), seen on its first HCON pass but not on the second six months later.  
However the Scanpi scans on the paired single HCON show it to have been detected on two out of four scans, so not a mover.

My triples search, requiring one association at 1-6 arcmin, and a second at 10-35 arcmin, yields 75 triplets, all of which I examined with 
Scanpi and in NED.  One of these, 314.83499+64.21527, is definitely of interest as a candidate for a mover and is discussed in the next section.

Finally a search for close pairs of single HCONs, with separations 2-6 arcmin, yielded 76 pairs (after excluding Saturn glints).  These were 
examined in Scanpi without yielding any further cases of interest. 

\begin{table*}
\caption{Unidentified IRAS sources with a missing HCON and with a single HCON partner at separation 1-30 arc min}
\begin{tabular}{lllllllll}
\hline
name & RA (1950) & dec &  b & S25 & S60 & S100 &  & comment \\
& (1950) &&& (Jy) &&& dist(`) & \\
\hline
{\bf PSCz} &&&&&&&&\\
Q05414+2128 & 85.358749 & 21.475834 & &  0.27 & 1.04 & 2.79 &  & undetected on 3rd HCON, close to predicted Planet 9 orbit\\
R05419+2156 & 85.48292 & 21.93889 & &  0.326 & 0.631 & 2.490 & 28.6 & none of 7 scans show a convincing detections \\ 
&&&&&&&&\\
{\bf RIIFSCz} &&&&&&&&\\
F20189-0408 & 304.745880 & -4.145520 & &  $<$0.310 & 0.321 & $<$1.736 &  & \\
R20195-0430 & 304.88586  & -4.50111 & &  $<$0.250 & 0.419 & 1.304 & 22.9 & 2M gal at 9 arcsec, zsp=0.1106, not unidentified \\
&&&&&&&&\\
F19499-2111 & 297.479980 & -21.197281 & &  $<$0.195 & 0.340 & $<$0.894 &  & faint sources, br star at 6 arcsec\\
R19515-2105 & 297.87750 & -21.09750 & & $<$0.458 & 0.565 & $<$1.000 & 23.0 & br 2M gal at 36 arcsec, zsp=0.03360, not unidentified \\
&&&&&&&&\\
F18264-6330 & 276.61043 & -63.50469 & &  $<$0.113 & 0.497 & $<$1.666 &  & not really undetected on 3rd HCON\\
R18242-6346 & 276.06082 & -63.78194 & &  $<$0.250 & 0.925 & $<$1.437 & 22.4 & undetected on first two HCONs but poor flux match \\
&&&&&&&&\\
{\bf QIGC} &&&&&&&&\\
738014-235334 & 114.50584 & -23.89278 & &  $<$0.25 & 0.55 & $<$2.17 &  & \\
R07378-2349 & 114.46250 & -23.82944 & &  $<$0.604 & 0.603 & $<$2.641 & 4.5 & non-detection on first two HCONs unconvincing \\
&&&&&&&&\\
1827362+223803 & 276.90085 & 22.63417 & &  $<$0.25 & 0.56 & $<$1.43 &  & \\
R18272+2246 & 276.81000 & 22.77472 & &  $<$0.250 & 0.448 & $<$1.274 & 9.8 & radio, br 2M gal at 9 arcsec, not undetected\\
&&&&&&&&\\
{\bf 1HCONs} &&&&&&&&\\
R18212+0719 & 275.30018 & 7.33028 & & $<$0.275 & 0.548 & $<$1.671 &  & non-detection on 2nd HCON\\
R18216+0724 & 275.42291 & 7.41306 & &  $<$0.250 & 0.513 & $<$2.480 & 9.0 & source seen on 2 out of 4 scans on 1st HCON \\
&&&&&&&&\\
R20593+6413 & 314.83499 & 64.21527 & &  $<$0.250 & 0.558 & (2.509) &  & clear non-detection on 3rd and 4th HCON\\
R20592+6411 & 314.80624 & 64.18917 & &  $<$0.250 & 0.627 & $<$3.251 & 1.7 & clear non-detection on 3rd and 4th HCON\\
R20562+6408 & 314.07373 & 64.14889 & &  $<$0.250 & 0.538 & $<$2.680 & 20.3 & clear non-detection on first 3 HCONs, {\bf Planet 9 candidate}\\

\hline
\end{tabular}
\end{table*}

\begin{figure*}
\includegraphics[width=7.0cm]{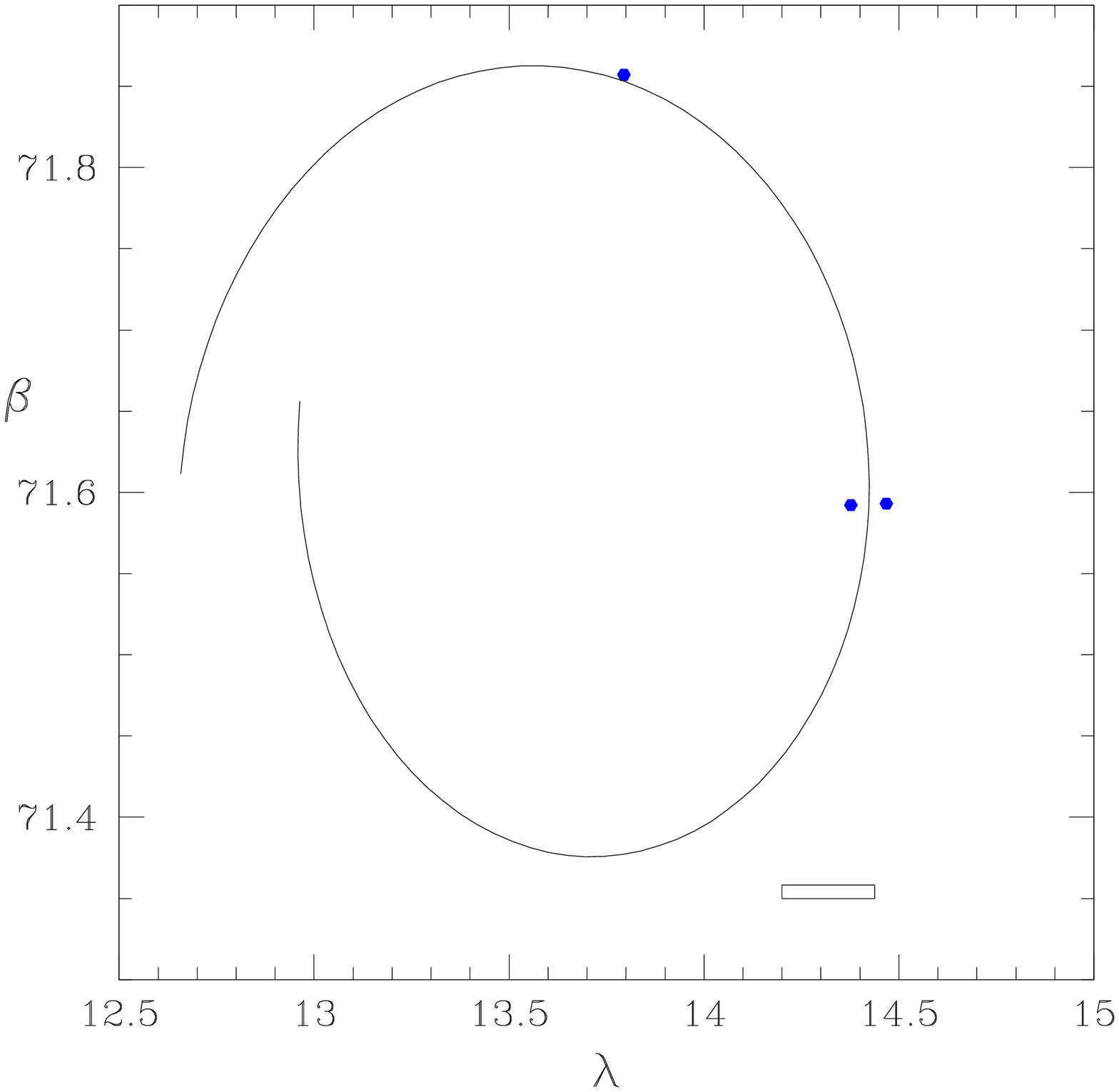}
\includegraphics[width=7.0cm]{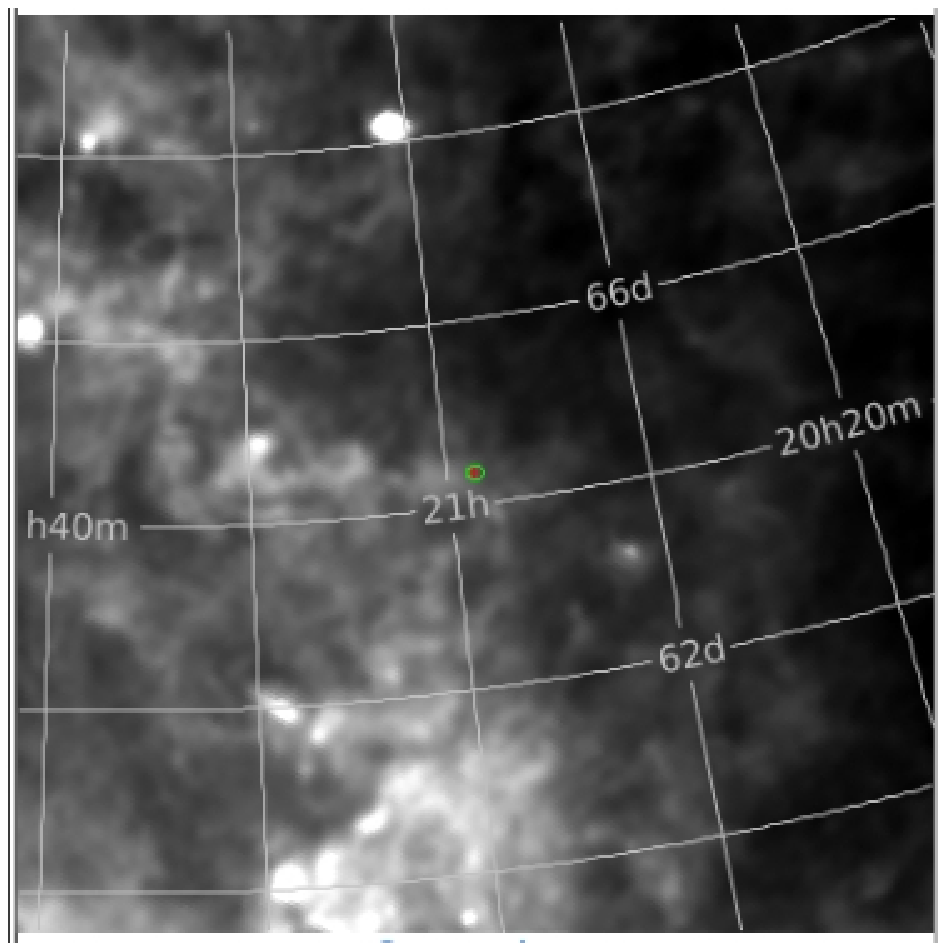}
\caption{
L: Possible orbit for planet 9 candidate for the year 1983 (ecliptic coordinates 2000).  The footprint of an IRAS 60 $\mu$m detector is indicated at the lower right.
R: Schlegel, Finkbeiner and Davis (1998) extinction map centred on location of Planet 9 candidate.  The estimated $A_V$ at this location is $>$2 mag.
}
\end{figure*}

\begin{figure*}
\includegraphics[width=14cm]{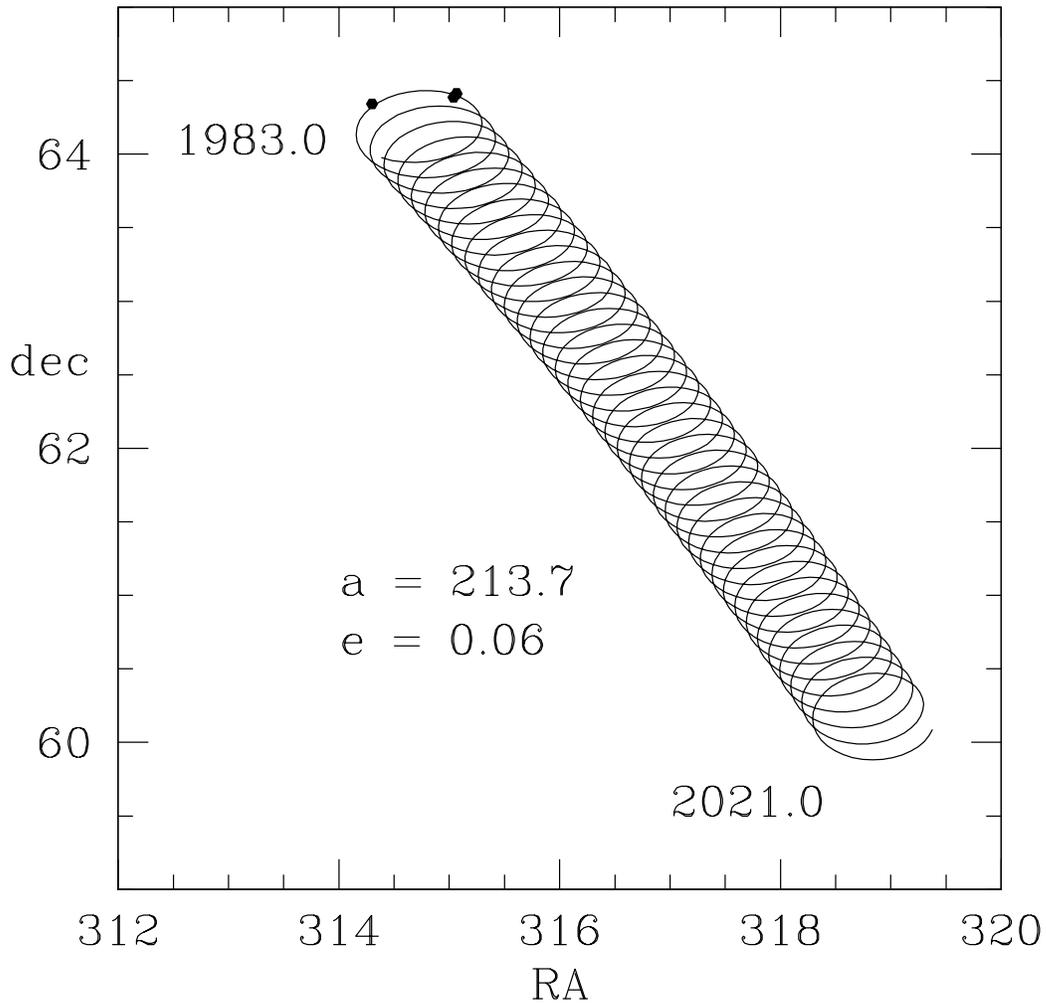}
\caption{
Possible orbit of candidate planet 9 (equatorial coordinates 2000) from 1983 to 2100. Filled circles show IRAS detections. }
\end{figure*}


\subsection{A planet 9 candidate}
The triplet of single HCONs R20593+6413, R20592+6411, R20562+6408, is definitely of interest as a candidate for 
Planet 9.  There were in fact four hours-confirming passes over this area of sky: (1) June 27-28, 1983, (2) July 13, 
(3) July 23-24 (4) Sept 19-20.  The Scanpi detection profile 
of the 3 single HCONs during these passes is shown in Table 2, which gives the ratio of the number of detections 
$>2.5\sigma/$number of possible detections.

HCONs 1 and 2, separated by 1.7 arcmin, appear to be cross-scan split 
detections of the same source, detected most strongly on pass 1.
Neither were detected on passes 3 or 4.  HCON 3 was detected on pass
4 but not on passes 1-3.  So the pattern of detections and non-detections
matches that expected for a slow-moving object, which has moved 20.3 arcmin in 11.9 weeks,
or 1.7 arcmin per week. 

Before we consider what the orbit of such an object might be, it should be noted that
the source(s) lie at low Galactic latitude (b = 9 deg), in a region strongly affected
by cirrus (Fig 7R).  The detections are not of high quality, do not show a strong correlation
with a point-source profile, and therefore allow a suspicion of being extended and thus cirrus.

\begin{table}
\caption{Detection profile of the 3 single HCONs during the 4 HCON passes}
\begin{tabular}{llllll}
\hline
pass & (1) & (2) & (3) & (4)\\
&&&&\\
R20593+6413 & 4$/$5 & 2$/4$ & 0$/$2 & 0$/$7\\
&&&&\\
R20592+6411 & 6$/$6  & 3$/$5 & 0$/$4 & 0$/$5\\
&&&&\\ 
R20562+6408 & 0$/$4 & 0$/$6 & 0$/$2 & 4$/$7\\	
\hline
\end{tabular}
\end{table}

The expected orbit of Planet 9 is a combination of a parallactic ellipse with major axis parallel to the
ecliptic plane, semi-major axis $\alpha (arcmin) = 3438/d(AU)$ and semi-minor axis $\alpha cos(\beta)$,
and the proper motion, which we may approximate as a straight line of length $d^{-1.5}$ per year, with
position angle (in ecliptic coordinates) $\theta$.

This very approximate orbit fit gives d = 225 $\pm$ 15AU, $\theta = 205 \pm 80^o$, and centre of the parallactic ellipse
at ($\lambda_0, \beta_0$)(2000.0) = (13.62, 71.62) $\pm 1^o$, and is illustrated in Figure 7L.  Scanpi signal-to-noise 
values of all the possible detections at 
the locations on the preferred orbit corresponding to the times of the four HCON passes are shown in Table 3.
For a slow-moving source on this orbit we should ideally see good detections on the diagonal of this array and none off the diagonal.
While the outcome is not ideal, it is quite suggestive.  This is even more suggestive in the plots of individual scans (available online), 
where 7 of the 8 off-diagonal scans 
do not show any evidence of a source at the central location on the scan, while all of the on-diagonal ones do.
Despite the poorer quality of the IRAS data at this low Galactic latitude, both the HCON data and the individual averaged detector scans
are consistent with the idea of the detection of a moving object.
  
Figure 8 shows a possible orbit from 1983 to 2100, calculated using The Project Pluto package Find Orb.  However the IRAS data are not accurate 
enough to allow an orbit to be determined and we can only say that at 2100.0 the candidate object should be within an 
annulus of radius 2.5-4$^o$, centred on ($\lambda_0, \beta_0$)(2000.0) = (13.62, 71.62).

\begin{table*}
\caption{Scanpi signal-to-noise at the locations on the preferred orbit corresponding to the four HCON passes}
\begin{tabular}{llllllll}
\hline
time	& $\lambda$(2000) & $\beta$ &	scan (1) 		&		scan (2)		&	scan (3)	& scan (4)\\
\hline
1983 June 28 6.7hrs & 14.4214 & 71.5869 & 2.9, {\bf 3.1}, 2.47, 0, 2.7, 2.8 & 2.7, 0, 2.7, 2.2, 2.47 & 0, 2.1, 0 & 0, 0, 0, 0\\
1983 July 13 17.6hrs & 14.4063 & 71.6551 & 2.9 2.1 2.4 2.6 2.46 2.8 & 0, {\bf 3.1}, 0, 2.49 & {\bf 3.0}, 0 & {\bf 3.6}, 0, 0, 0, 0, {\bf 3.8}, 0\\
1983 July 23 23.4hrs & 14.3722 & 71.6913 & 0, 0, 0, 0, 0, 2.49 & 0, 0, 0, {\bf 3.1}, 0 & 2.9, 0 & 0, 0, 0, 0, 0 \\
1983 Sept 20 5.7hrs & 13.7554 & 71.8560 & 0,0, 2.2, 0 & 0, 0, 0, 0, 0, 0 & 0, 0, 0, 0 & {\bf 3.1}, 2.4, 2.3, 2.7, {\bf 3.4}, 2.6\\
\hline
\end{tabular}
\end{table*}

Figure 9 shows planetary temperature versus radius, with the expected locus of a 60$\mu$m source of 0.57 Jy (red loci, corresponding to distances of 210, 225 and 240 AU) together 
with locations of the Fortney et al (2019) Planet 9 models. The strong cirrus in the vicinity of the source
make a significant 100$\mu$m detection or upper limit impossible.  

The candidate discussed here would be likely to have mass 3-5 $M_E$,
with the ephemerides limits (Batygin et al 2019) suggesting a mass at the lower end of the range. However this candidate is rather different from the Batygin and Brown (2019) proposal on which the ephemerides limit of Fienga et al (2016), from which the Batygin et al (2019) constraint is extrapolated, is based. This candidate is of significantly lower mass and at much higher ecliptic latitude.  It would clearly be desirable to reconsider the ephemerides limits for an object with this orbit.

\begin{figure}
\includegraphics[width=9cm]{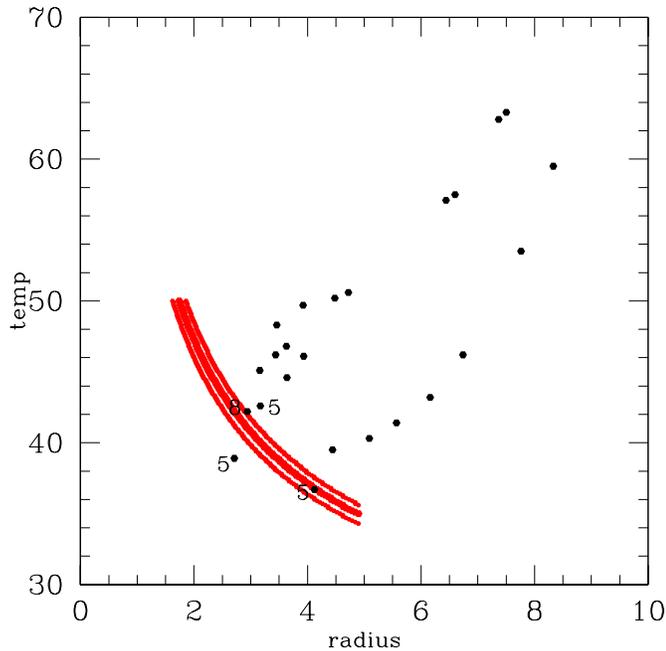}
\caption{
Planetary temperature versus radius ($r_E$) for Fortney et al (2019) planetary models (filled hexagrams).  
Red curves: loci on which S60 = 0.57 Jy blackbody source would lie, for distance 210 (lower locus), 225, 240 AU.}
\end{figure}

\section{Recovery of Planet 9 candidate 1983-present}
A planet of mass 3-5 $M_E$, radius 1.5-5 $R_E$, at a distance of 210-240 AU, should in principle be relatively
easy to recover from survey data in the intervening period. The major difficulty is the closeness of the candidate 
planet to the Galactic Plane, making the detection of even a relatively bright moving object difficult against 
the very high surface density of stars.

The Palomar Digital Sky Survey (DPOSS) surveyed the entire northern sky in the late 1980s and early 1990s and released
a catalogue but only for $|b| > 10^0$ (Reid et al 1991, Djorgovski 1999).

Meissner et al (2018) carried out a search for moving objects in the WISE survey data, but only for 70$\%$ of sky, and not including the area of interest.

The best dataset for this search is the Panstarrs ‘3 $\pi$’ survey in bands g,r,i,z,y of the entire sky north of dec = -30$^o$, carried 
out in 2009-14 (Chambers et al 2019). Approximately half of the gri observations were carried out in a series of 4 exposures, quads, all separated
by about 15 minutes, so completed within 1 hour.  By analogy with IRAS I will refer to these sets of observations taken within about an hour as HCONs.  
These observations were then repeated several times each year over the 5 years of the survey.  In one hour our Planet 9 candidate would have moved 
about 0.6 arcsec, so would be registered as a point-source, ie a single HCON, albeit with only 4 detections out of the target 60 for the survey.  
Wainscott et al (2015)  reported  a search for
Near Earth Objects (NEOs) with Panstarrs, but not yet at high ecliptic latitude or low Galactic latitude.

Mallama et al (2017) give predicted colours and magnitudes for solar system planets, and we can use these to estimate
colours and magnitudes for Neptune, Jupiter or Pluto -like objects, scaled for r=3.66 $R_E$
at a distance 225 AU , with an uncertainty of $\pm$ 0.3 dex in distance and radius. For Jupiter-like colours, (g,r,i,z = 16.84, 16.26, 16.41, 16.73),
for Neptune-like colours, (g, r, i, z = 16.79, 16.95, 18.04, 18.90), and for Pluto-like colours, (g, r, i, z = 18.02. 17.22, 16.99, 16.99), respectively. 
These suggest the search can be restricted to (r-i) $<$ 1.0 and (g-r) $<$ 1.5.  



I used VizieR to extract Panstarrs single HCONs within 2.5-4.0 deg of the 1983 position.  Many are too bright to be of interest, so I restricted the search to $i>14$.
The majority are part of point sources seen on all 4 or 5 years of observation, so I restricted the search to sources with less than 5 detections.  I examined detections 
for each candidate using MAST.  There were a small number of genuine single HCONs but none form the expected radial line of HCONs corresponding to the 
5 years of Panstarrs observation. Moreover none of the ‘single HCONs’ in this area of the sky are in fact true quads.  Typically one triplet of gri observations 
was carried out on a single night, within one hour, while other repeat observations are carried out 4-10 nights later.
Since in 4-10 days our candidate object would have moved $\sim$ 1-3 arcmin, different bands would be displaced by this amount and would not be combined into a
single source. However Planet 9 could still have been seen as sets
of triplet detections in a roughly radial line, a year apart in time.  A search for HCON triplets in the expected annulus did not result 
in any candidates of interest.

The failure to recover the candidate in the Panstarrs data suggests that this candidate object is not in fact real, though the fact that the quad strategy was not implemented in the region of sky of interest makes this conclusion uncertain.  Given the poor quality of the IRAS detections, at the very limit of the survey, and in a very difficult part of the sky for far infrared detections, the probability of the candidate being real is not overwhelming. However given the
great interest of the Planet 9 hypothesis, it would be worthwhile to check whether an object with the proposed parameters and in the region of sky proposed, is inconsistent with the planetary ephemerides.  Fienga et al (2020) state that a 5 $M_E$ planet would have to be at a distance $>$ 500 AU to be consistent with the planetary ephemerides, but their simulations do not appear to include cases at such high ecliptic latitudes as my candidate.  If simulations showed that the candidate here is not inconsistent with the planetary ephemerides, a more sophisticated search of the Panstarrs database would be worthwhile. 



\section{Discussion}
The analysis here takes the discussion of identifications of the PSCz galaxy catalogue a little further than Saunders et al (2000), and
the identification of RIIFSCz sources significantly further than Wang et al (2014).
0.3$\%$ of the PSCz galaxies at $|b|>5^o$ (52 sources) remain unidentified,  
and 3$\%$ of RIIFSCz sources remain unidentified.

I have given a fairly full discussion of the nature of the IRAS 60 $\mu$m single HCON sources in the IRAS PSC Reject Catalogue.  
Most are main belt asteroids, galaxies or cirrus.
I do not find a convincing example of an unidentified PSCz or RIIFSCz source associated with an unidentified single HCON at a separation of 1-30
arc min, which would be a candidate for a ninth planet. Iorio (2014, 2016) and Fienga et al (2016, 2020) have already given dynamical arguments 
which severely constrain the type of object postulated by Batygin and Brown (2016).

Searching for close pairs or triplets of single HCONs, I have found a candidate for a $\sim 3-5 M_E$ planet at $\sim$ 210-240 AU in the 1983 IRAS data. It is at very high ecliptic latitude
and at low Galactic latitude, which may explain why it has not been found in previous searches for a 9th
planet. Dynamical studies are needed to check whether such an object is consistent with the ephemerides of
other solar system objects and whether this object can account for the clustering of the orbits of Kuiper
belt dwarf planets.
The IRAS detections are not of the highest quality but it would be worth searching at optical and near infrared wavelengths
in an annulus of radius 2.5-4 deg centred on the 1983 position.  This candidate could be ruled out if radio
or other observations confirmed the reality (and stationarity) of the IRAS sources at the 1983 1HCON positions.

IRAS had a significant capability to detect a planet in the outer solar system of mass 10 $M_E$ or greater, especially if it
had an extended gaseous envelope.  For the 70$\%$ of the sky which had a third 
IRAS HCON coverage, the region of Fig 2 to the left of the 60 $\mu$m limit 
is excluded.  However even within the areas that did not receive a third HCON, Planet 9 would have been detectable
as an unidentified PSCz source (for d $>$ 400 AU) or as a close pair of single HCONs (for d =200-400 AU).  
IRAS would have been able to detect a 10 $M_E$ planet to 380-600 AU, and could have detected a 20 $M_E$
planet to 690-1000 AU (with the higher limit corresponding to an extended planetary atmosphere).  The 90$\%$ completeness limit 
for these searches at 60 $\mu$m is 0.45 Jy for $|b|>20^0$.  Between $|b|=20^0$ and the IRAS mask (see Fig 5), the completeness 
limit is 90$\%$ down to 0.6 Jy, and 80$\%$ between 0.5 and 0.6 Jy.  These are worthwhile constraints for the Planet 9 search.
 
\section{Acknowledgements}
This work would not have been possible without the data tools and date archives provided by IRSA, NED, HEASARC, VizieR and MAST.

\section {Data availability} 
All data used in this paper have been taken from public archives.

\section{Note added in proof}
In a paper published after this paper was submitted, Brown and Batygin (2021, AJ in press) have carried out new simulations of Planet 9
candidates capable of explaining the clustering of orbits of Kuiper belt dwarf planets.  Their preferred planetary mass, 6.2 +2.2 -1.3 $M_E$, and perihelion,
300 +85 -60 AU, are now quite close to those of my Planet 9 candidate.  However the preferred direction on the sky for their Planet 9 object is very far from my candidate, so my object, if real, may not be the cause of the orbit clustering.


\begin{thebibliography}{99}

\bibitem{} Batygin K., Brown M.E., 2016, AJ 151, 22
\bibitem{} Batygin K., Adams F.C., Brown M.E., Becker J.C., 2019, Physics Reports 805, 1
\bibitem{} Brown M.E., Batygin K., 2016, ApJ 824, L23
\bibitem{} Brown M.E., Batygin K., 2019, AJ 157, 62
\bibitem{} Caceres J., Gomes R., 2018, AJ 156, 157
\bibitem{} Crosswell K., 1991, New Scientist, Nov 30
\bibitem{} Davies J.K. et al, 1984, Nature 309, 315
\bibitem{} Fienga A., Laskar J., Manche H., Gastineau M., 2016, AA 587, L8
\bibitem{} Fienga A., Di Ruscio A., Bernus L., Deram P., Durante D., Laskar J., Iess L., 2020, AA 640, A6
\bibitem{} Fortney J.J., Marley M.S., Laughlin G. et al, 2016, ApJL 824, L25
\bibitem{} Harrington R.S., 1988, AJ 96, 1476
\bibitem{} Holman M.J., Payne M.J., 2016, AJ 152, 94
\bibitem{} Iorio L., 2017, Astrophys.Space.Sci.Supp. 362, 111
\bibitem{} IRAS Explanatory Supplement, 1988, eds C.A.Beichmann, G.Neugebauer, H.J.Habing, P.E.Clegg, Chester, T.J. (NASA$/$STI)
\bibitem{} The IRAS Asteroid and Comet Survey, 1986, ed. D.L.Matson, JPL D-3698 (Pasadena: JPL)
\bibitem{} Jackson A.A., Killen R.M., 1988, MNRAS 235, 593
\bibitem{} Kraan-Korteweg R.C., Lahav O., 2000, Astron.Astroph.Rev. 10, 211, 
\bibitem{} Luhman K.L.,2014, ApJ 781, 4
\bibitem{} Mallama A., Krobusek B., Pavlov H., 2017, Icarus 282, 19
\bibitem{} Matthews R., 1991 Science 254, 1454 
\bibitem{} Medvedev Y.D., Vavilov D.E., Bondarenko Y.S., Bulekbaev D.A., Kunturova N.B., 2017, Astr.L. 43, 120
\bibitem{} Meisner A.M., Bromley B.C., Nugent P.E., et al, 2017 AJ 153, 65
\bibitem{} Meisner A.M., Bromley B.C., Kenyon S.J., Andersin T.E., 2018, AJ 155, 166
\bibitem{} Millholland S., Laughton G., 2017, AJ 153, 91
\bibitem{} Mordasini C. Alibert Y., Georgy C., Dittkrist K.-M., Klahr H., Henning T., 2012, AA 547, 112
\bibitem{} Morrison L.V., 1992, Observatory 112, 36
\bibitem{} Moshir M. et al, 1992,Explanatory Supplement to IRAS Faint Source Survey, Version 2. JPL D-100015 8/92 (Pasadena JPL)
\bibitem{} Murray J.B., 1999, MNRAS 309, 31
\bibitem{} Neuhauser R., Feitzinger J.V., 1991, Earth, Moon and Planets, 54, 193
\bibitem{} Reynolds R.T. et al, 1980, Icarus 44, 772
\bibitem{} Rowan-Robinson M., Saunders W., Lawrence, Leach K., 1991, MNRAS 253, 485
\bibitem{} Rowan-Robinson M., 2013, {\it Night Vision}, CUP
\bibitem{} Saunders W. et al,2000, MNRAS 317, 55 
\bibitem{} Schlegel D.J., Finkbeiner D.P., Davis M., 1998, ApJ 500, 525
\bibitem{} Seidelmann P.K. , Harrington R.S., 1988, Celestial Mechanics 43, 55
\bibitem{} Standish M., 1992, AJ 105, 200
\bibitem{} Staveley-Smith L. et al, 2016, AJ 151, 52
\bibitem{} Trujillo C.A., Sheppard S.S., 2014, Nature 507, 471
\bibitem{} Wainscott R., Chambers K., Lilly E., Weryk R., Chastel S., Denneau L., Micheli M., 2015,in IAU
Symposium 318, Asteroids: New observations, New Models, eds S.R.Chesley, A.Morbidelli, R.Jedicki, D.Farnocchia. p.293
\bibitem{} Walker R.G., Rowan-Robinson M., 1984, BAAS 16, 443
\bibitem{} Wang L., Rowan-Robinson M., 2010, MNRAS 398, 109
\bibitem{} Wang L., Rowan-Robinson M., Norberg P., Heinis S., Han J., 2014, MNRAS 442, 2739





\end{thebibliography}
\end{document}